%% file: ANA-TOPQ-2026-07-PAPER.tex
\documentclass[PAPER,UKenglish,cernpreprint,atlasdraft=false, texmf, orcidlogo]{atlasdoc}
\usepackage{atlaspackage}
\usepackage{atlasbiblatex}

\usepackage{atlasphysics}
\graphicspath{{logos/}{figures/}}

\usepackage{ANA-TOPQ-2026-07-PAPER-defs}

\AtlasTitle{Study of $t\bar{t}$ threshold effects in $e\mu$ differential distributions measured in $\boldmath\sqrt{s}=13\,\TeV$ $\boldmath pp$ collisions with the ATLAS detector
}

\AtlasAbstract{%
The recent ATLAS measurement of the normalised $e\mu$ dilepton invariant mass
and azimuthal angle distributions in $\sqrt{s}=13$\,\TeV\ $pp$ collisions at
the Large Hadron Collider is extended to study the sensitivity
to the formation of quasi-bound states near the \ttbar\ threshold.
The measured differential distributions are compared with \ttbar\ models
incorporating perturbative QCD predictions for the hard process, with or
without the addition of colour-singlet quasi-bound states. The
data are better described by the predictions incorporating quasi-bound-state
formation. Fits to the dilepton invariant mass distribution show
evidence for the latter process with an observed significance exceeding
three standard deviations, and a measured cross-section in agreement with
dedicated studies of the threshold region.
}

\AtlasRefCode{TOPQ-2026-07}

\PreprintIdNumber{CERN-EP-2026-107}

\AtlasJournal{EPJC}
\hypersetup{pdftitle={ATLAS document},pdfauthor={The ATLAS Collaboration}}

\begin{document}

\maketitle

\tableofcontents

\section{Introduction}

The study of top-quark--antiquark (\ttbar) production at the CERN Large Hadron
Collider (LHC) allows quantum chromodynamics (QCD) to be probed at some of the
highest accessible energy scales, and forms a key part of the LHC experimental
programme. Recently, the modelling of the \ttbar\ threshold region, where the
\ttbar\ invariant mass \mtt\ is close to twice the top quark mass of
$\mtop\approx 172.5$\,\GeV, has received particular attention.
Using combinations of \mtt\ and angular variables, both
CMS~\cite{CMS-TOP-24-007} and ATLAS~\cite{TOPQ-2025-11} have reported excesses
at low \mtt\ compared with the predictions of standard \ttbar\ event generators
based on a perturbative QCD (pQCD) description of the hard-scattering process,
and that appear compatible with the formation of a quasi-bound state
close to threshold.

Precise measurements of lepton and dilepton kinematic distributions in \ttbar\
events provide important information for developing and tuning event generator
models of \ttbar\ production. The recent ATLAS measurement of lepton differential
cross-sections in $\ttbar\rightarrow e\mu\nu\bar{\nu}\bbbar$ events from
140\,\ifb\ of $pp$ collisions at \sxyt ~\cite{TOPQ-2024-12} provides a
wide range of precise differential measurements including the dilepton invariant
mass \mll\ and dilepton azimuthal angular difference \dphill, as well as the
measurement of \dphill\ in coarse bins of \mll. The invariant mass distribution
shows a significant data excess in the region $\mll<100$\,\GeV\ when compared
with pQCD predictions from \ttbar\ event generators.
This excess was also present in previous
measurements of \mll\ with a 36\,\ifb\ dataset at \sxyt~\cite{TOPQ-2018-17}, and
at \sxvt~\cite{TOPQ-2015-02}. The \mll, \dphill\
and two-dimensional \dphillvmll\ distributions
measured in Ref.~\cite{TOPQ-2018-17}, along with related measurements
from CMS~\cite{CMS-TOP-17-014}, were fitted to a combination of pQCD and
quasi-bound-state predictions in Ref.~\cite{oniastat},
showing that the data can be better described
by predictions including the latter contribution, and claiming evidence for
the formation of toponium. These studies were performed
using the HEPData~\cite{hepdata} records of the measured distributions,
and did not take into account bin-to-bin correlations in the ATLAS data.

This paper follows a similar approach, comparing the distributions measured
in Ref.~\cite{TOPQ-2024-12} with various predictions with or without
quasi-bound-state contributions, normalised according to a theoretical
prediction. In a second step, the \mll, \dphill\ and
\dphillvmll\ distributions are also fitted to predictions where the normalisation
of the quasi-bound-state contribution is treated as a free parameter. The
analysis exploits the precision of the full Run~2 dataset recorded by the ATLAS
detector~~\cite{PERF-2007-01,PIX-2018-001,ATLAS-TDR-19,ATL-SOFT-PUB-2021-001},
with a larger data sample and smaller lepton efficiency and \ttbar\ modelling
uncertainties than in the 36\,\ifb\ measurement~\cite{TOPQ-2018-17}.
It also takes into account bin-to-bin
correlations in the uncertainties in both the experimental measurements and
the predictions. The predictions are described in Section~\ref{s:mcmod}
and the analysis methodology in Section~\ref{s:meth}. Results are given
in Section~\ref{s:res} and conclusions drawn in Section~\ref{s:conc}.

\section{Simulated event samples}\label{s:mcmod}

The measured distributions were compared with particle-level predictions in the
fiducial region defined by the \ttem\  measurement~\cite{TOPQ-2024-12},
namely an opposite-charge
$e\mu$ pair, with each lepton satisfying $\pt>20$\,\GeV\ and $|\eta|<2.5$, with
no requirements on jets.\footnote{ATLAS uses a right-handed coordinate system
with its origin at
the nominal interaction point in the centre of the detector, and the $z$ axis
along the beam line. Pseudorapidity is defined in terms of the polar angle
$\theta$ as $\eta=-\ln\tan{\theta/2}$, and transverse momentum and energy
are defined relative to the beam line as $\pt=p\sin\theta$ and
$\et=E\sin\theta$. The azimuthal angle around the beam line is denoted by
$\phi$.
The rapidity is defined as $y=\frac{1}{2}\ln\left(\frac{E+p_z}{E-p_z}\right)$,
where $p_z$ is the $z$-component of the momentum and $E$ is the energy
of the relevant object or system.} For the pQCD \ttbar\ predictions covering
the inclusive phase space, the following three models were considered,
all of which are described in more detail in Ref.~\cite{TOPQ-2024-12}:
\begin{description}
\item[{\scr Powheg\,+\,Pythia8}:] The QCD next-to-leading-order (NLO) matrix
element for \ttbar\ production implemented in the {\ttr hvq} process in
{\scr Powheg Box} v2~\cite{powheghvq}, interfaced to
{\scr Pythia8}~\cite{pythia82}.
\item[{\scr Powheg\,+\,Pythia8}] NNLO rew: The {\scr Powheg\,+\,Pythia8} sample
described above, reweighted to bring the top quark kinematics into agreement
with the predictions of the next-to-next-to-leading-order (NNLO)
QCD\,+\,NLO electroweak calculation
of Ref.~\cite{ttnnlocalc} as described in Ref.~\cite{TOPQ-2020-14}.
The reweighting was performed iteratively in the three variables
$p_\mathrm{T}(t)$, \mtt\ and $p_\mathrm{T}(\ttbar)$ to ensure a good description
of all three distributions.
This reweighting differs from that used in the
dedicated ATLAS \ttbar\ threshold study~\cite{TOPQ-2025-11}, which uses
a different set of variables.
\item[{\scr Powheg MiNNLO\,+\,Pythia8}:] The prediction obtained from the
\minnlops\ matched NNLO QCD plus parton shower (PS)
procedure~\cite{powhegminnlo},
implemented in {\scr Powheg}~\cite{powheg,powheg2,powheg3} and interfaced
to {\scr Pythia8}.
\end{description}
The latter two predictions give a better overall description of the lepton
and dilepton kinematics than the first prediction from
{\scr Powheg\,+\,Pythia8}. For comparison with the \embb\ distributions
measured in Ref.~\cite{TOPQ-2024-12}, which include both \ttbar\ and $Wt$
contributions, and also make fiducial requirements
on the presence of two $b$-tagged jets with $\pt>25$\,\GeV\ and $|\eta|<2.5$,
the \powhegbbfl\ sample was also considered. This sample was generated
with the {\scr Powheg} $bb4l$ generator~\cite{jezobb4l}, which
includes the complete set of $\ell\ell'\nu\bar{\nu}\bbbar$
final states, taking into account interference between the \ttbar\
and $Wt$ final states as well as off-shell and non-resonant effects.
All these samples use the NNPDF3.0 parton distribution function (PDF)
set~\cite{nnpdf3} in the matrix element calculation, the {\scr Pythia8}
A14 tune~\cite{ATL-PHYS-PUB-2014-021}, {\scr EvtGen}~\cite{evtgen}
to handle the decay of $b$- and $c$-flavoured hadrons,
and assume $\mtop=172.5$\,\GeV.

The quasi-bound-state effects were modelled using the two approaches
described in Ref.~\cite{TOPQ-2025-11}, in each case adding to the pQCD
\ttbar\ predictions discussed above. The first approach describes the
formation of a quasi-bound colour-singlet \ttbar\ state using the Green's
function of non-relativistic QCD (NRQCD) in the Coulomb gauge~\cite{topofuks}.
The results were used to reweight the production and
decay of a six-body final state generated at leading order (LO) in QCD with
{\scr MadGraph 3.4.2}~\cite{madgraph} and the NNPDF3.0 PDF set, interfaced to
{\scr Pythia8}.
The inclusive production cross-section was set to 6.43\,pb, as obtained from
analytic calculations~\cite{toponiumxsec}. This model is referred to as
Green's function reweighting or
\ttonia. The second approach is that originally used in the
CMS publication~\cite{CMS-TOP-24-007}, i.e.\ a simplified model with a
pseudo-scalar resonance \etatt\ of mass 343\,\GeV\ and width 2.8\,\GeV.
The resonance was simulated
using {\scr MadGraph 3.5.5} at LO to generate the four-body $WbWb$ final state
that was then interfaced to {\scr Pythia8}. The same cross-section as
for the \ttonia\ approach was assumed.

\section{Methodology}\label{s:meth}

The consistency of each prediction with a measured normalised differential
cross-section was quantified using a $\chi^2$ calculated as
\begin{eqnarray}\label{e:redchi}
\chi^2 = {\mathbf\Delta}_{(n-1)}^T {\mathbf S}^{-1}_{(n-1)} {\mathbf\Delta}_{(n-1)}\,,
\end{eqnarray}
where ${\mathbf\Delta}_{(n-1)}$ is the vector of differences between the measured
and predicted normalised differential cross-section in each of the $n$
bins, excluding the last one, and ${\mathbf S}^{-1}_{(n-1)}$ is the inverse of
the corresponding
covariance matrix, including both the experimental uncertainties in the
measurement and the uncertainties in the predictions, and their
bin-to-bin correlations through off-diagonal terms. The last bin of each
distribution was excluded to account for the normalisation condition.\footnote{
The $\chi^2$ does not depend on which bin is excluded.}
The uncertainties in the pQCD predictions were calculated as discussed in
Ref.~\cite{TOPQ-2024-12}, and include PDF, QCD renormalisation and factorisation
scales, initial/final state radiation (IFSR) and top quark mass variations.
The NNPDF3.0 replica sets were used to calculate the bin-to-bin correlations
for the PDF uncertainty, whilst the QCD scale, radiation and top quark mass
uncertainties were each assumed to be fully correlated between bins.
For predictions
that include a component from quasi-bound-state effects, added with a fixed
normalisation corresponding to
the assumed cross-section of 6.43\,pb, the same relative uncertainties as in
the pQCD-alone predictions were assumed, and applied to the total prediction.

In a second interpretation, the quasi-bound-state prediction was added to the
pQCD prediction with a normalisation given by a signal strength parameter
$\mu$, where $\mu=1$ corresponds to the predicted cross-section of 6.43\,pb.
The $\chi^2$ of Eq.~(\ref{e:redchi}) was then minimised to determine the
best-fit value of $\mu$ and its uncertainty. These fits were performed
for the \mll, \dphill\ and \dphillvmll distributions only.
Separate uncertainties for PDF, QCD renormalisation and factorisation
scales, IFSR and the use of {\scr Herwig7}~\cite{herwigpp,herwig71}
rather than {\scr Pythia8} for parton shower
and hadronisation were calculated for the quasi-bound-state contribution, and
considered independent of the uncertainties in the pQCD component.

\section{Results}\label{s:res}

The $\chi^2$ of Eq.~(\ref{e:redchi}) was calculated for each pQCD model
alone, and for each model with the addition of the \ttonia\ or \etatt\
quasi-bound-state model. The $\chi^2$ values and associated probabilities
for each of the ten one-dimensional
distributions measured in Ref.~\cite{TOPQ-2024-12} are shown in
Table~\ref{t:reschioned}, and for the three two-dimensional distributions
of \etal, \rapll\ and \dphill\ in bins of \mll\ in Table~\ref{t:reschitwod}.
The comparisons for predictions involving  \powhegbbfl\ are based on
the measured \embb\ distributions, and the others on the \ttem\ distributions.
The values for the pQCD-alone predictions are the same as those in
Ref.~\cite{TOPQ-2024-12}.

\begin{table}[tp]
\caption{\label{t:reschioned}$\chi^2$ values (top block) and associated
probabilities
(bottom block) for comparison of normalised measured single-differential
fiducial cross-sections with various predictions without and with
quasi-bound-state contributions simulated with the \ttonia\ and \etatt\ models.
Within each block, the last three rows correspond to comparisons
of the \embb\ cross-sections with combined \ttbar+$Wt$ samples, and the
other rows to \ttem\ cross-sections compared with \ttbar\ samples.}
\footnotesize

\hspace{-10mm}
\begin{tabular}{lrrrrrrrrrr}\toprule
Distribution & \ptl & \etal & \ptll & \mll & \rapll & \dphill & \ptsum & \esum & \ptlmax & \ptlmin \\
$N_\mathrm{dof}$ & 12 & 8 & 10 & 14 & 8 & 19 & 9 & 11 & 11 & 10 \\ \midrule
$\chi^2$ values & \\ \cmidrule{1-1}{\scr Powheg\,+\,PY8}&  35.2 &  19.7 &  12.5 &  49.8 &  11.6 &  22.7 &  34.9 &  14.7 &  32.0 &  27.4 \\
{\scr Powheg\,+\,PY8}\,+\,\ttonia&  30.1 &  20.0 &  11.1 &  29.0 &  11.6 &  19.4 &  31.4 &  11.6 &  26.4 &  24.9 \\
{\scr Powheg\,+\,PY8}\,+\,\etatt&  30.6 &  20.1 &  11.1 &  30.0 &  11.6 &  19.4 &  31.6 &  12.0 &  26.9 &  25.1 \\
{\scr Powheg\,+\,PY8} NNLO rew&  16.4 &  14.9 &  11.7 &  33.6 &  12.8 &  16.5 &  24.6 &  11.3 &  14.3 &  14.6 \\
{\scr Powheg\,+\,PY8} NNLO rew\,+\,\ttonia&  16.7 &  15.1 &  15.4 &  23.8 &  13.0 &  18.8 &  24.6 &  10.4 &  14.1 &  14.2 \\
{\scr Powheg\,+\,PY8} NNLO rew\,+\,\etatt&  16.6 &  15.2 &  15.3 &  23.8 &  13.0 &  18.5 &  24.6 &  10.6 &  14.1 &  14.4 \\
{\scr Powheg MiNNLO\,+\,PY8}&  15.4 &  12.3 &   7.2 &  25.9 &  11.7 &  23.6 &  20.9 &   7.5 &  12.1 &  12.5 \\
{\scr Powheg MiNNLO\,+\,PY8}\,+\,\ttonia&  14.4 &  12.4 &   7.3 &  16.5 &  11.9 &  16.7 &  20.7 &   6.6 &  11.0 &  12.0 \\
{\scr Powheg MiNNLO\,+\,PY8}\,+\,\etatt&  14.4 &  12.5 &   7.3 &  16.5 &  11.9 &  17.2 &  20.6 &   6.7 &  10.9 &  12.0 \\
\cmidrule{1-1}{\scr Powheg} $bb4l$\,+\,{\scr PY8}&  15.5 &  11.2 &   3.1 &  20.5 &  14.7 &  21.4 &  20.5 &  18.6 &  19.2 &  21.8 \\
{\scr Powheg} $bb4l$\,+\,{\scr PY8}\,+\,\ttonia&  14.8 &  11.6 &   4.1 &  14.1 &  14.7 &  19.9 &  18.6 &  15.6 &  18.3 &  19.4 \\
{\scr Powheg} $bb4l$\,+\,{\scr PY8}\,+\,\etatt&  14.6 &  11.6 &   4.1 &  14.0 &  14.7 &  19.9 &  18.8 &  15.9 &  18.2 &  19.5 \\
\midrule[\heavyrulewidth]
$\chi^2$ probabilities & \\
\cmidrule{1-1}{\scr Powheg\,+\,PY8}& $4\ 10^{-4}$ & 0.011 & 0.25 & $7\ 10^{-6}$ & 0.17 & 0.25 & $6\ 10^{-5}$ & 0.20 & $8\ 10^{-4}$ & $2\ 10^{-3}$ \\
{\scr Powheg\,+\,PY8}\,+\,\ttonia& $3\ 10^{-3}$ & 0.010 & 0.35 & 0.011 & 0.17 & 0.43 & $3\ 10^{-4}$ & 0.39 & $6\ 10^{-3}$ & $6\ 10^{-3}$ \\
{\scr Powheg\,+\,PY8}\,+\,\etatt& $2\ 10^{-3}$ & 0.010 & 0.35 & $8\ 10^{-3}$ & 0.17 & 0.43 & $2\ 10^{-4}$ & 0.37 & $5\ 10^{-3}$ & $5\ 10^{-3}$ \\
{\scr Powheg\,+\,PY8} NNLO rew& 0.18 & 0.061 & 0.31 & $2\ 10^{-3}$ & 0.12 & 0.62 & $3\ 10^{-3}$ & 0.41 & 0.22 & 0.15 \\
{\scr Powheg\,+\,PY8} NNLO rew\,+\,\ttonia& 0.16 & 0.057 & 0.12 & 0.048 & 0.11 & 0.47 & $4\ 10^{-3}$ & 0.49 & 0.23 & 0.16 \\
{\scr Powheg\,+\,PY8} NNLO rew\,+\,\etatt& 0.17 & 0.056 & 0.12 & 0.049 & 0.11 & 0.49 & $3\ 10^{-3}$ & 0.47 & 0.23 & 0.16 \\
{\scr Powheg MiNNLO\,+\,PY8}& 0.22 & 0.14 & 0.71 & 0.026 & 0.16 & 0.21 & 0.013 & 0.76 & 0.35 & 0.25 \\
{\scr Powheg MiNNLO\,+\,PY8}\,+\,\ttonia& 0.28 & 0.13 & 0.70 & 0.28 & 0.16 & 0.61 & 0.014 & 0.83 & 0.44 & 0.29 \\
{\scr Powheg MiNNLO\,+\,PY8}\,+\,\etatt& 0.28 & 0.13 & 0.70 & 0.28 & 0.16 & 0.58 & 0.014 & 0.83 & 0.45 & 0.28 \\
\cmidrule{1-1}{\scr Powheg} $bb4l$\,+\,{\scr PY8}& 0.22 & 0.19 & 0.98 & 0.11 & 0.064 & 0.32 & 0.015 & 0.069 & 0.057 & 0.016 \\
{\scr Powheg} $bb4l$\,+\,{\scr PY8}\,+\,\ttonia& 0.25 & 0.17 & 0.94 & 0.45 & 0.065 & 0.40 & 0.029 & 0.16 & 0.076 & 0.035 \\
{\scr Powheg} $bb4l$\,+\,{\scr PY8}\,+\,\etatt& 0.27 & 0.17 & 0.94 & 0.45 & 0.065 & 0.40 & 0.027 & 0.14 & 0.078 & 0.034 \\
\bottomrule
\end{tabular}
\end{table}

\begin{table}[tp]
\caption{\label{t:reschitwod}$\chi^2$ values (top block) and associated
probabilities
(bottom block) for comparison of normalised measured double-differential
fiducial cross-sections with various simulation
samples. Within each block, the last three rows correspond to comparisons
of the \embb\ cross-sections with combined \ttbar+$Wt$ samples, and the
other rows to \ttem\ cross-sections compared with \ttbar\ samples.}
\small\centering

\begin{tabular}{lrrr}\toprule
Distribution & \etalvmll & \rapllvmll & \dphillvmll \\
$N_\mathrm{dof}$ & 35 & 19 & 39 \\ \midrule
$\chi^2$ values & \\ \cmidrule{1-1}{\scr Powheg\,+\,PY8}&  72.7 &  46.0 &  79.9 \\
{\scr Powheg\,+\,PY8}\,+\,\ttonia&  60.6 &  33.7 &  63.2 \\
{\scr Powheg\,+\,PY8}\,+\,\etatt&  61.0 &  34.4 &  64.2 \\
{\scr Powheg\,+\,PY8} NNLO rew&  71.0 &  33.8 &  71.0 \\
{\scr Powheg\,+\,PY8} NNLO rew\,+\,\ttonia&  64.4 &  27.1 &  71.2 \\
{\scr Powheg\,+\,PY8} NNLO rew\,+\,\etatt&  64.6 &  27.4 &  71.3 \\
{\scr Powheg MiNNLO\,+\,PY8}&  44.8 &  26.7 &  50.2 \\
{\scr Powheg MiNNLO\,+\,PY8}\,+\,\ttonia&  38.1 &  21.4 &  44.3 \\
{\scr Powheg MiNNLO\,+\,PY8}\,+\,\etatt&  37.9 &  21.3 &  44.0 \\
\cmidrule{1-1}{\scr Powheg} $bb4l$\,+\,{\scr PY8}&  47.8 &  23.7 &  66.0 \\
{\scr Powheg} $bb4l$\,+\,{\scr PY8}\,+\,\ttonia&  45.6 &  18.0 &  65.9 \\
{\scr Powheg} $bb4l$\,+\,{\scr PY8}\,+\,\etatt&  45.3 &  18.1 &  65.9 \\
\midrule[\heavyrulewidth]
$\chi^2$ probabilities & \\
\cmidrule{1-1}{\scr Powheg\,+\,PY8}& $2\ 10^{-4}$ & $5\ 10^{-4}$ & $1\ 10^{-4}$ \\
{\scr Powheg\,+\,PY8}\,+\,\ttonia& $5\ 10^{-3}$ & 0.020 & $8\ 10^{-3}$ \\
{\scr Powheg\,+\,PY8}\,+\,\etatt& $4\ 10^{-3}$ & 0.017 & $7\ 10^{-3}$ \\
{\scr Powheg\,+\,PY8} NNLO rew& $3\ 10^{-4}$ & 0.020 & $1\ 10^{-3}$ \\
{\scr Powheg\,+\,PY8} NNLO rew\,+\,\ttonia& $2\ 10^{-3}$ & 0.10 & $1\ 10^{-3}$ \\
{\scr Powheg\,+\,PY8} NNLO rew\,+\,\etatt& $2\ 10^{-3}$ & 0.096 & $1\ 10^{-3}$ \\
{\scr Powheg MiNNLO\,+\,PY8}& 0.12 & 0.11 & 0.11 \\
{\scr Powheg MiNNLO\,+\,PY8}\,+\,\ttonia& 0.33 & 0.32 & 0.26 \\
{\scr Powheg MiNNLO\,+\,PY8}\,+\,\etatt& 0.34 & 0.32 & 0.27 \\
\cmidrule{1-1}{\scr Powheg} $bb4l$\,+\,{\scr PY8}& 0.073 & 0.21 & $4\ 10^{-3}$ \\
{\scr Powheg} $bb4l$\,+\,{\scr PY8}\,+\,\ttonia& 0.11 & 0.52 & $4\ 10^{-3}$ \\
{\scr Powheg} $bb4l$\,+\,{\scr PY8}\,+\,\etatt& 0.11 & 0.51 & $5\ 10^{-3}$ \\
\bottomrule
\end{tabular}
\end{table}

The measured absolute \ttem\ and \embb\ differential
cross-sections as functions of \mll\ and \dphill\ are shown in
Figure~\ref{f:distresa}, together with panels showing the ratio of each
normalised prediction to data, demonstrating the effect of adding the \ttonia\
or \etatt\ prediction to each pQCD \ttbar\ or $\ttbar+Wt$ model.
The two-dimensional \dphillvmll\ distribution and corresponding ratios
are shown in Figure~\ref{f:distresb}.

\begin{figure}[tp]
\parbox{83mm}{\includegraphics[width=76mm]{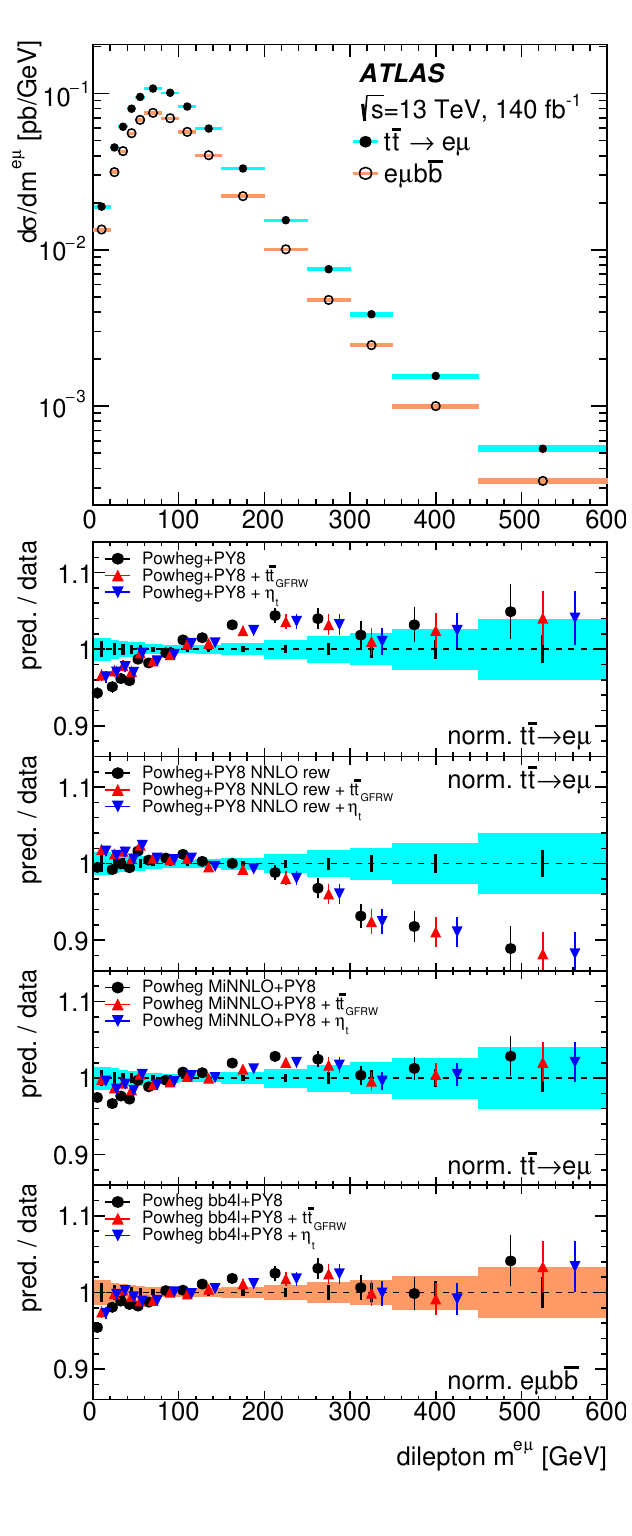}\vspace{-7mm}\center{(a)}}
\parbox{83mm}{\includegraphics[width=76mm]{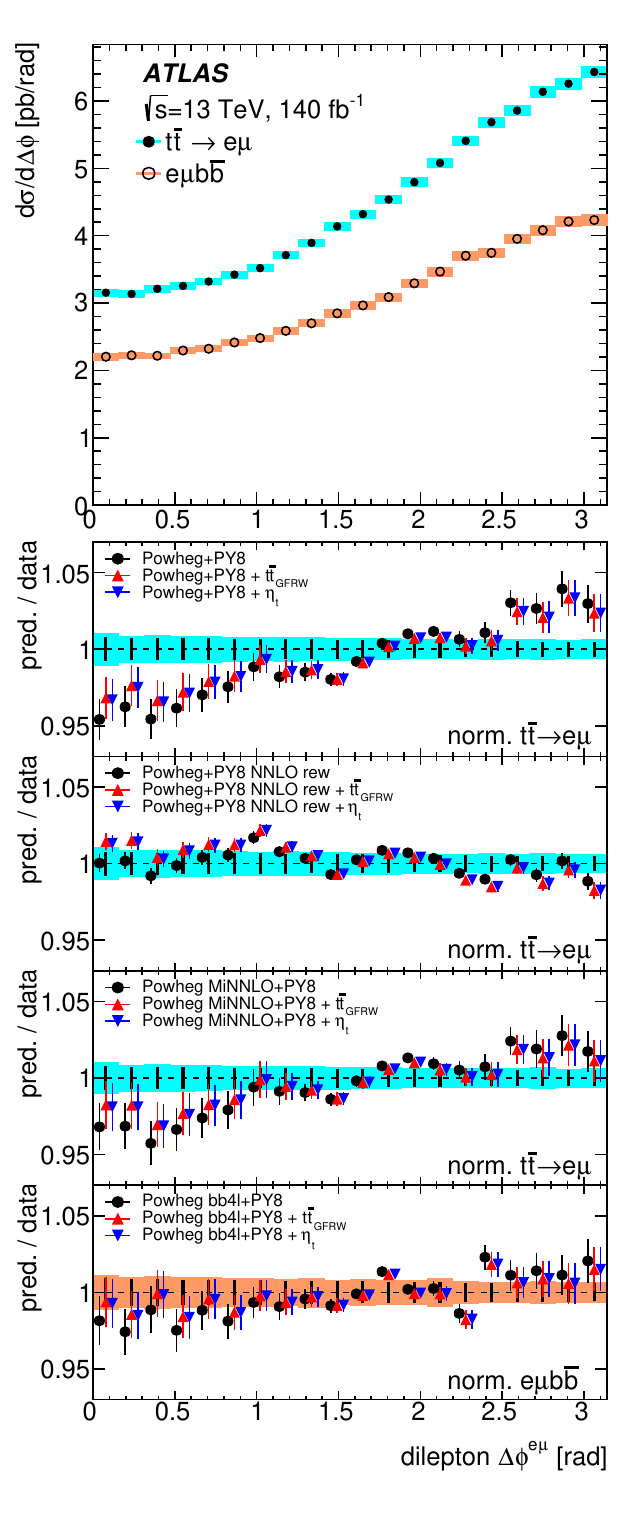}\vspace{-7mm}\center{(b)}}
\caption{\label{f:distresa}Differential cross-sections as functions of (a) \mll\
and (b) \dphill. The upper panels show the measured absolute \ttem\ (filled points)
and \embb\ (open points) cross-sections, including the contributions from
$W\rightarrow\tau\rightarrow e/\mu$ decays, with the total uncertainties shown
by the shaded bands. The other panels show ratios of various predictions
without and with quasi-bound-state contributions simulated with the \ttonia\
and \etatt\ models to the
measured normalised differential cross-sections, for \ttem\ (middle three panels)
and \embb\ (lower panels). The markers showing the ratios for each prediction
are offset from the bin centres for better visibility, and the total
uncertainty in the prediction is shown by the error bar. The data statistical
uncertainty is shown by the black error bars and the total uncertainty by the
cyan (\ttem) or orange (\embb) band around unity.}
\end{figure}

\begin{figure}[tp]
\centering

\includegraphics[width=160mm]{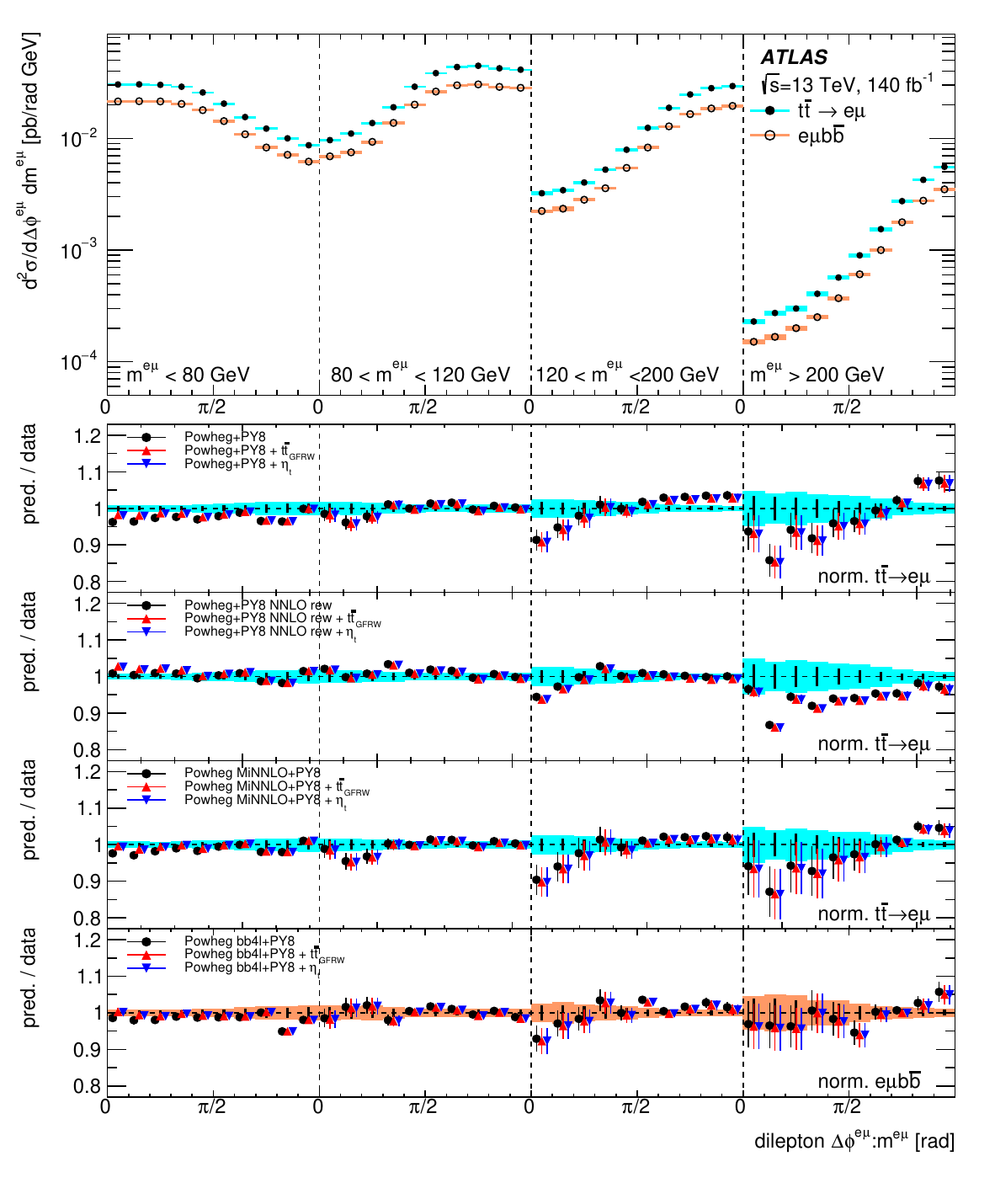}
\caption{\label{f:distresb}Double-differential cross-section as a function
of \dphill\ and \mll, showing the unrolled cross-sections as a function of
\dphill\ in four bins of \mll. The upper panel shows the measured absolute \ttem\
(filled points)
and \embb\ (open points) cross-sections, including the contributions from
$W\rightarrow\tau\rightarrow e/\mu$ decays, with the total uncertainties shown
by the shaded bands. The other panels show ratios of various predictions
without and with quasi-bound-state contributions simulated with the \ttonia\
and \etatt\ models to the
measured normalised differential cross-sections, for \ttem\ (middle three panels)
and \embb\ (lower panel). The markers showing the ratios for each prediction
are offset from the bin centres for better visibility, and the total
uncertainty in the prediction is shown by the error bar. The data statistical
uncertainty is shown by the black error bars and the total uncertainty by the
cyan (\ttem) or orange (\embb) band around unity.}
\end{figure}

The modelling of the \mll\ distribution is significantly improved by the
addition of the \ttonia\ quasi-bound-state component, reflected both
in the $\chi^2$ values and ratio plots. The \ttonia\ contribution adds
about 3\% to the absolute differential cross-section for $\mll<30$\,\GeV,
reducing to a negligible contribution for $\mll>100$\,\GeV.
This addition improves the $\chi^2$ from 25.9 to 16.5 for 14~degrees of freedom.
The NNLO-reweighted {\scr Powheg\,+\,Pythia8} pQCD prediction shows a significant
deficit with respect to the data at high \mll, which has the effect of reducing
the discrepancy in the normalised distribution at low \mll. However, the
addition of the \ttonia\ contribution still improves the overall agreement
and $\chi^2$ values, in this case from~33.6 to~23.8.
The \mll\ distribution of the \etatt\ model is very similar
to that of \ttonia, and the improvements in $\chi^2$ are very similar.

Smaller improvements are seen for the \dphill\ and \dphillvmll\ distributions
when adding the \ttonia\ or \etatt\ contributions to the pQCD \ttbar\
predictions. In the case of the NNLO-reweighted {\scr Powheg\,+\,Pythia8}
prediction, where the modelling of \dphill\ is already good,
adding \ttonia\ slightly increases the $\chi^2$ value. The first \mll\ bin in
the two-dimensional distributions covers $\mll<80$\,\GeV, covering
the entire range where the quasi-bound-state contribution is significant,
and where it varies rapidly as a function of \mll. At low \mll, \dphill\ and
\mll\ are also correlated, due to the requirement that both leptons have
$\pt>20$\,\GeV, which strongly restricts the selected phase space.
The finer bins of the
one-dimensional \mll\ distribution therefore provide greater sensitivity
than the two-dimensional \dphillvmll\ distribution with the present binning.

Tables~\ref{t:reschioned} and~\ref{t:reschitwod} also show the $\chi^2$ values
for all the other one- and two-dimensional distributions measured in
Ref.~\cite{TOPQ-2024-12}. For these distributions, the changes in $\chi^2$
from adding the quasi-bound-state contributions are not significant,
demonstrating that these distributions do not provide sensitivity to
\ttbar\ threshold effects.

The results of fitting the quasi-bound-state signal strengths $\mu$ using each
of the \mll, \dphill\ and \dphillvmll\ distributions
are shown numerically in Table~\ref{t:oniafit} and
graphically in Figure~\ref{f:oniafit}.
Predictions using the \powhegminnlo\ and NNLO-reweighted
{\scr Powheg\,+\,Pythia8} pQCD \ttbar\ models were fitted together with
the \ttonia\ and
\etatt\ quasi-bound state models using the \ttem\ results, and
\powhegbbfl\ was fitted together with each quasi-bound-state model using
the \embb\ results.

\begin{table}
\caption{\label{t:oniafit}Summary of the fits for quasi-bound-state
contributions with floating normalisation. The columns show the combination
of pQCD and quasi-bound-state models, the fitted normalised  distribution,
the number of degrees of freedom, the $\chi^2$ at the minimum and the
corresponding fitted value of $\mu$ with its total uncertainty
including statistical, experimental and prediction systematic contributions.
The \ttem\ distributions are fitted for the first four groups, and the
\embb\ distributions for the last two.}
\centering
\begin{tabular}{llrrc}\toprule
Model & Distribution & $N_\mathrm{dof}$ & $\chi^2_\mathrm{min}$ & $\mu$ \\ \midrule{\scr Powheg MiNNLO\,+\,PY8} + \ttonia & \mll & 13 & 16.2 & $1.22^{+0.38}_{-0.38}$\rule[-1mm]{0mm}{6mm} \\
& \dphill & 18 & 14.5 & $2.10^{+0.95}_{-0.78}$\rule[-1mm]{0mm}{6mm} \\
& \dphillvmll & 38 & 43.2 & $0.81^{+0.32}_{-0.31}$\rule[-1mm]{0mm}{6mm} \\
\cmidrule{1-1}{\scr Powheg\,+\,PY8} NNLO rew + \ttonia & \mll & 13 & 23.8 & $1.02^{+0.34}_{-0.33}$\rule[-1mm]{0mm}{6mm} \\
& \dphill & 18 & 16.3 & $0.18^{+0.51}_{-0.48}$\rule[-1mm]{0mm}{6mm} \\
& \dphillvmll & 38 & 67.4 & $0.62^{+0.34}_{-0.33}$\rule[-1mm]{0mm}{6mm} \\
\cmidrule{1-1}{\scr Powheg MiNNLO\,+\,PY8} + \etatt & \mll & 13 & 15.8 & $1.38^{+0.40}_{-0.41}$\rule[-1mm]{0mm}{6mm} \\
& \dphill & 18 & 14.2 & $2.45^{+0.93}_{-0.85}$\rule[-1mm]{0mm}{6mm} \\
& \dphillvmll & 38 & 43.4 & $0.86^{+0.33}_{-0.33}$\rule[-1mm]{0mm}{6mm} \\
\cmidrule{1-1}{\scr Powheg\,+\,PY8} NNLO rew + \etatt & \mll & 13 & 23.6 & $1.14^{+0.37}_{-0.37}$\rule[-1mm]{0mm}{6mm} \\
& \dphill & 18 & 16.4 & $0.19^{+0.54}_{-0.52}$\rule[-1mm]{0mm}{6mm} \\
& \dphillvmll & 38 & 68.4 & $0.54^{+0.35}_{-0.34}$\rule[-1mm]{0mm}{6mm} \\
\midrule
{\scr Powheg} $bb4l$\,+\,{\scr PY8} + \ttonia & \mll & 13 & 14.0 & $1.19^{+0.44}_{-0.45}$\rule[-1mm]{0mm}{6mm} \\
& \dphill & 18 & 19.9 & $1.25^{+1.04}_{-1.02}$\rule[-1mm]{0mm}{6mm} \\
& \dphillvmll & 38 & 63.9 & $0.60^{+0.43}_{-0.42}$\rule[-1mm]{0mm}{6mm} \\
\cmidrule{1-1}{\scr Powheg} $bb4l$\,+\,{\scr PY8} + \etatt & \mll & 13 & 13.7 & $1.34^{+0.47}_{-0.49}$\rule[-1mm]{0mm}{6mm} \\
& \dphill & 18 & 19.8 & $1.44^{+1.10}_{-1.13}$\rule[-1mm]{0mm}{6mm} \\
& \dphillvmll & 38 & 64.3 & $0.58^{+0.44}_{-0.44}$\rule[-1mm]{0mm}{6mm} \\
\bottomrule
\end{tabular}
\end{table}

\begin{figure}[tp]
\parbox{83mm}{\includegraphics[width=83mm]{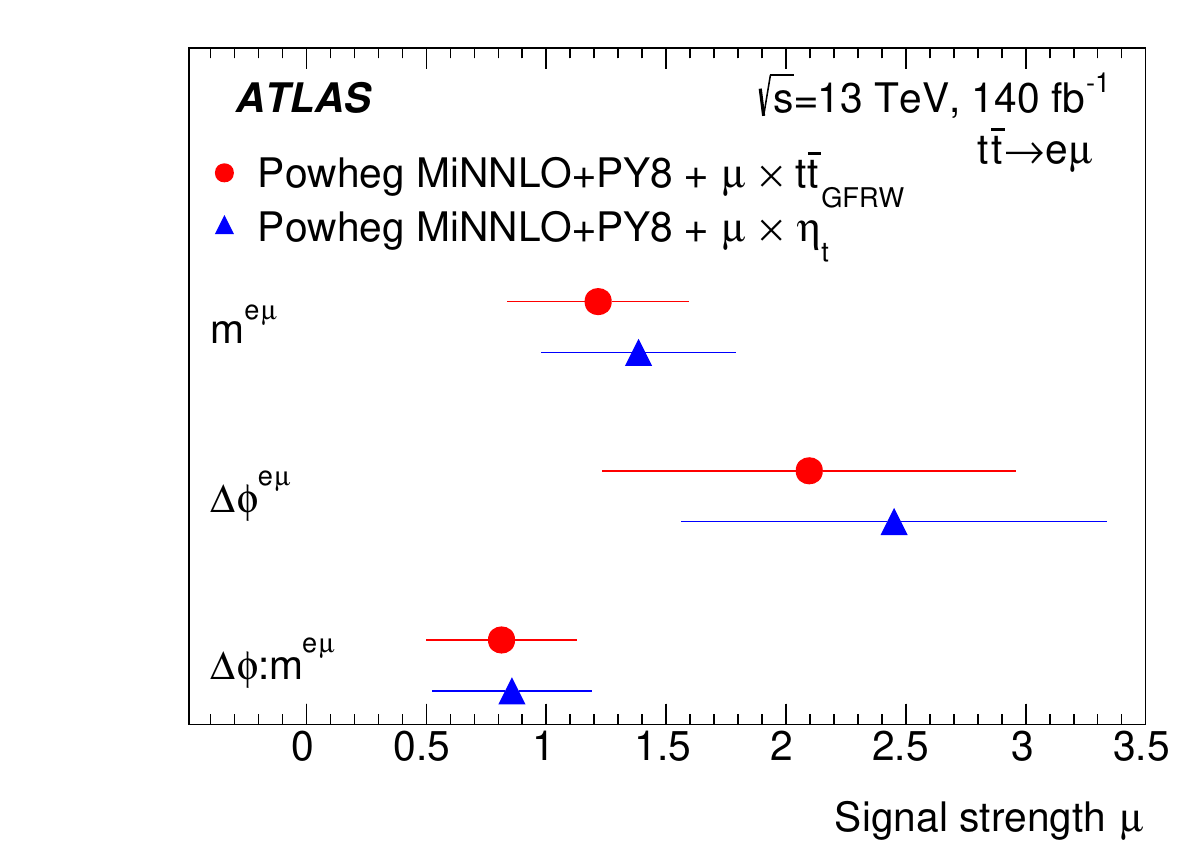}\vspace{-7mm}\center{(a)}}
\parbox{83mm}{\includegraphics[width=83mm]{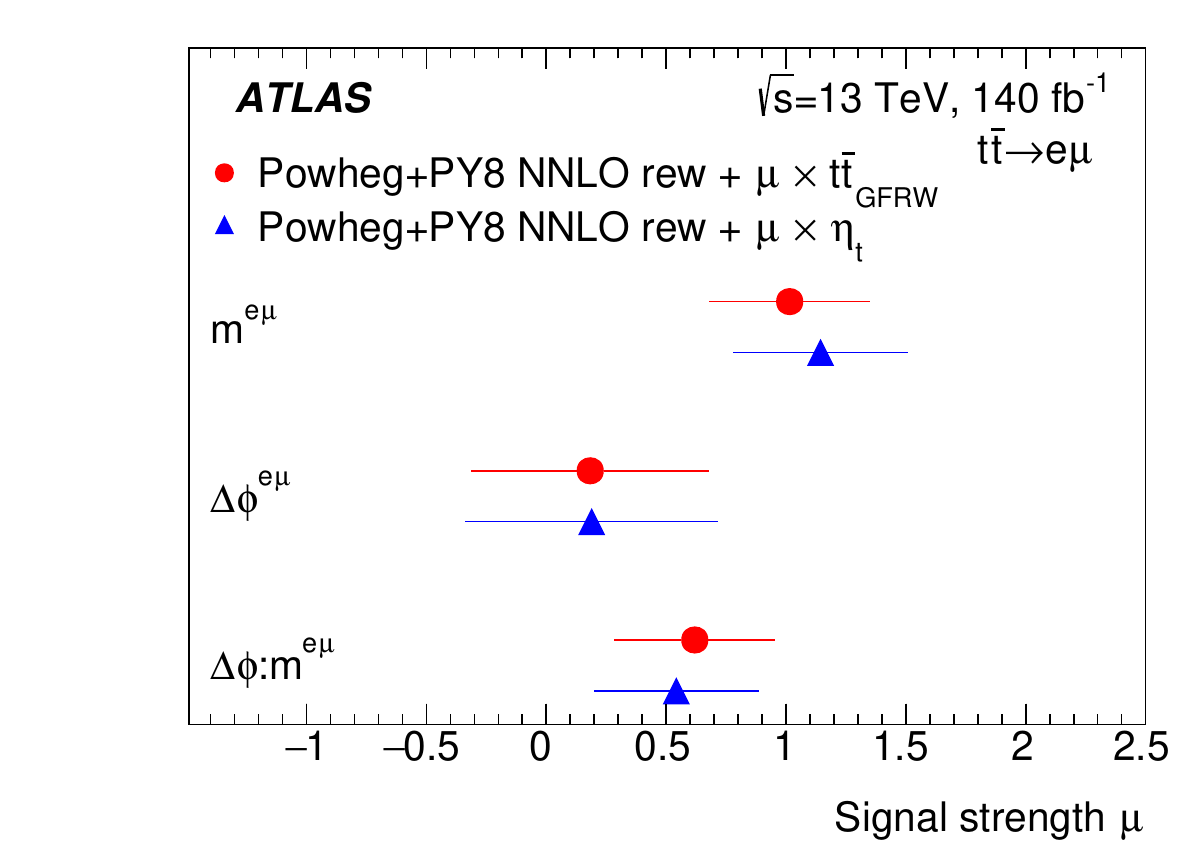}\vspace{-7mm}\center{(b)}}
\parbox{83mm}{\includegraphics[width=83mm]{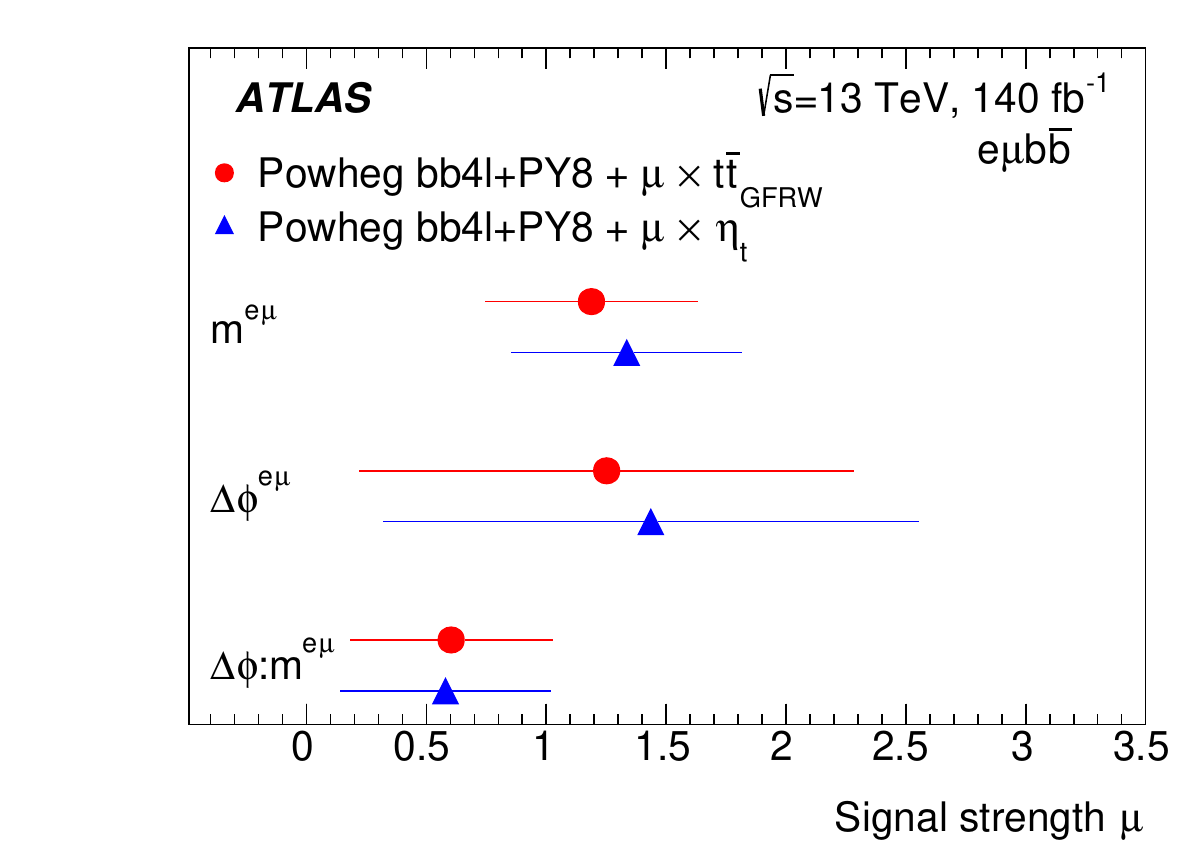}\vspace{-7mm}\center{(c)}}
\caption{\label{f:oniafit}Fitted $\mu$ values, the signal strength of the
quasi-bound-state contribution,
using \ttonia\ (red circles) and \etatt\ (blue triangles)
models fitted to each of the \mll, \dphill\ and \dphillvmll\ distributions,
with the pQCD contribution modelled using (a)~\powhegminnlo,
(b) NNLO-reweighted {\scr Powheg\,+\,Pythia8} and (c) \powhegbbfl. The
error bars show the full uncertainties from the fit, including statistical,
experimental systematic and prediction-related uncertainties.}
\end{figure}

The fits to the \ttem\ \mll\ distributions all give values of $\mu$ that
are compatible with one, and differ from zero by more than three standard
deviations. The fits to the \embb\ \mll\ distribution with \powhegbbfl\
as the pQCD $\ttbar+Wt$ prediction give $\mu$ values that are compatible with
those from the \ttem\ fits and
differ from zero by more than two standard deviations. The experimental
uncertainties in the \embb\ distributions are larger than those for \ttem,
due to the extra requirements on $b$-tagged jets and the associated
systematic uncertainties.
The fits to the \dphill\ distribution have larger uncertainties and
greater spread than those for \mll,
with the results generally favouring $\mu=1$, but also being compatible
with $\mu=0$ for the NNLO-reweighted {\scr Powheg\,+\,Pythia8} pQCD model
and both quasi-bound-state models. The results from the \dphillvmll\
distribution are compatible with $\mu=1$, but do not strongly exclude
zero, especially in the case of the NNLO-reweighted pQCD model.
In general, the largest contribution to the total uncertainty in $\mu$
comes from the experimental measurement, followed by the uncertainties
in the pQCD predictions. The results do not change significantly if the
QCD renormalisation and factorisation scale uncertainties are taken to be
correlated rather than uncorrelated.
The uncertainties in the quasi-bound-state model play only a minor role.

The best description of the \mll\ distribution in the \ttem\ selection
is given by the
\powhegminnlo\ pQCD \ttbar\ model combined with either the \ttonia\ or
\etatt\ quasi-bound-state model. Taking \ttonia\ as the more physically
motivated model, the signal strength is
\begin{equation}\nonumber
\mu = 1.22\pm 0.38
\end{equation}
corresponding to an observed cross-section of $7.8\pm 2.4$\,pb, close to
that observed in Ref.~\cite{TOPQ-2025-11} using a fit to angular
observables in fully reconstructed dileptonic \ttbar\ events.

\FloatBarrier

\section{Conclusion}\label{s:conc}

The sensitivity of leptonic differential distributions in
$\ttbar\rightarrow e\mu\nu\bar{\nu}\bbbar$ events to the formation of
quasi-bound states near the \ttbar\ threshold has been explored. Recent
ATLAS measurements of the \mll\ and \dphill\ distributions in $pp$ collisions
at \sxyt\ have been compared with the best-available \ttbar\ models based on
pQCD predictions,
with or without the addition of colour-singlet quasi-bound-state formation,
and taking bin-to-bin correlations into account.
The data are better described by predictions including the latter contribution,
modelled either using a non-relativistic QCD Green's function approach,
or a pseudo-scalar \etatt\ resonance. Fits to the \mll\ distribution for
the \ttem\ selection show
evidence for a quasi-bound-state contribution with a significance exceeding
three standard deviations, and a measured cross-section in agreement with
dedicated studies using the reconstructed \ttbar\ invariant mass distribution
and angular variables.

\section*{Acknowledgements}
%

%

%
%

%
%

We thank CERN for the very successful operation of the LHC and its injectors, as well as the support staff at
CERN and at our institutions worldwide without whom ATLAS could not be operated efficiently.

The crucial computing support from all WLCG partners is acknowledged gratefully, in particular from CERN, the ATLAS Tier-1 facilities at TRIUMF/SFU (Canada), NDGF (Denmark, Norway, Sweden), CC-IN2P3 (France), KIT/GridKA (Germany), INFN-CNAF (Italy), NL-T1 (Netherlands), PIC (Spain), RAL (UK) and BNL (USA), the Tier-2 facilities worldwide and large non-WLCG resource providers. Major contributors of computing resources are listed in Ref.~\cite{ATL-SOFT-PUB-2026-001}.

We gratefully acknowledge the support of ANPCyT, Argentina; YerPhI, Armenia; ARC, Australia; BMWFW and FWF, Austria; ANAS, Azerbaijan; CNPq and FAPESP, Brazil; NSERC, NRC and CFI, Canada; CERN; ANID, Chile; CAS, MOST and NSFC, China; Minciencias, Colombia; MEYS CR, Czech Republic; DNRF and DNSRC, Denmark; IN2P3-CNRS and CEA-DRF/IRFU, France; SRNSFG, Georgia; BMFTR, HGF and MPG, Germany; GSRI, Greece; RGC and Hong Kong SAR, China; ICHEP and Academy of Sciences and Humanities, Israel; INFN, Italy; MEXT and JSPS, Japan; CNRST, Morocco; NWO, Netherlands; RCN, Norway; MNiSW, Poland; FCT, Portugal; MNE/IFA, Romania; MSTDI, Serbia; MSSR, Slovakia; ARIS and MVZI, Slovenia; DSI/NRF, South Africa; MICIU/AEI, Spain; SRC and Wallenberg Foundation, Sweden; SERI, SNSF and Cantons of Bern and Geneva, Switzerland; NSTC, Taipei; TENMAK, T\"urkiye; STFC/UKRI, United Kingdom; DOE and NSF, United States of America.

Individual groups and members have received support from BCKDF, CANARIE, CRC and DRAC, Canada; CERN-CZ, FORTE and PRIMUS, Czech Republic; COST, ERC, ERDF, Horizon 2020 and Marie Sk{\l}odowska-Curie Actions, European Union; Investissements d'Avenir Labex, Investissements d'Avenir Idex and ANR, France; DFG and AvH Foundation, Germany; Herakleitos, Thales and Aristeia programmes co-financed by EU-ESF and the Greek NSRF, Greece; BSF-NSF and MINERVA, Israel; NCN and NAWA, Poland; La Caixa Banking Foundation, CERCA and AGAUR programs from Generalitat de Catalunya and PROMETEO and GenT Programmes Generalitat Valenciana, Spain; G\"{o}ran Gustafssons Stiftelse, Sweden; The Royal Society and Leverhulme Trust, United Kingdom; Eric and Wendy Schmidt Fund for Strategic Innovation, United States of America.

In addition, individual members wish to acknowledge support from Chile: Agencia Nacional de Investigaci\'on y Desarrollo (ANID FONDECYT reg. 1230987, FONDECYT 1230812, FONDECYT 1240864, Fondecyt 3240661, Fondecyt Regular 1240721); China: Chinese Ministry of Science and Technology (MOST-2023YFA1605700, MOST-2023YFA1609300), National Natural Science Foundation of China (NSFC 12275265, NSFC-W2543005); Czech Republic: Czech Science Foundation (GACR - 24-11373S), Ministry of Education Youth and Sports (ERC-CZ-LL2327, FORTE CZ.02.01.01/00/22\_008/0004632); EU: H2020 European Research Council (ERC - 101002463); European Union: European Research Council (BARD No. 101116429, ERC 101089007), European Regional Development Fund (HE COFUND GA No.101081355, ERDF), Marie Sklodowska-Curie Actions (GAP-101168829); France: Agence Nationale de la Recherche (ANR-21-CE31-0013, ANR-22-EDIR-0002, ANR-24-CE31-0504-01); Germany: Deutsche Forschungsgemeinschaft (DFG - 469666862); China: Research Grants Council (GRF); Italy: Ministero dell'Università e della Ricerca (NextGenEU I53D23000820006 M4C2.1.1, SOE2024\_0000023); Japan: Japan Society for the Promotion of Science (JSPS KAKENHI  JP25H0063, JSPS KAKENHI JP22H04944, JSPS KAKENHI JP24K23939, JSPS KAKENHI JP24KK0251, JSPS KAKENHI JP25H00650, JSPS KAKENHI JP25H01291, JSPS KAKENHI JP25K01011, JSPS KAKENHI JP25K01023, JSPS KAKENHI JP25KK0047); Poland: Polish National Science Centre (NCN 2021/42/E/ST2/00350, NCN OPUS 2023/51/B/ST2/02507, NCN OPUS nr 2022/47/B/ST2/03059, NCN UMO-2019/34/E/ST2/00393, UMO-2022/47/O/ST2/00148, UMO-2023/49/B/ST2/04085, UMO-2023/51/B/ST2/00920, UMO-2024/53/N/ST2/00869); Spain: Agència de Gestió d'Ajuts Universitaris i de Recerca. (AGAUR - 2023 BP 00141), Ministry of Science and Innovation (RYC2019-028510-I, RYC2020-030254-I, RYC2021-031273-I, RYC2022-038164-I), Ministerio de Ciencia, Innovación y Universidades/Agencia Estatal de Investigaci\'on (EU NextGenerationEU (PRTR-C17.I1), PID2022-142604OB-C22); Sweden: Carl Trygger Foundation (Carl Trygger Foundation CTS 22:2312), Swedish Research Council (Swedish Research Council 2023-04654, VR 2021-03651, VR 2022-03845, VR 2022-04683, VR 2023-03403, VR 2024-05451, VR 2025-05940), Knut and Alice Wallenberg Foundation (KAW 2023.0366); United Kingdom: The Binks Trust, Royal Society (NIF-R1-231091); United States of America: U.S. Department of Energy (ECA DE-AC02-76SF00515), John Templeton Foundation (John Templeton Foundation 63206), Neubauer Family Foundation.

%
%


%
\clearpage
\printbibliography
\clearpage
\input{atlas_authlist}
%

%
%
%
%

%
%
%
%

\end{document}

%% file: atlas_authlist.tex
 
\begin{flushleft}
\hypersetup{urlcolor=black}
{\Large The ATLAS Collaboration}

\bigskip

\AtlasOrcid[0000-0002-6665-4934]{G.~Aad}$^\textrm{\scriptsize 102}$,
\AtlasOrcid[0000-0001-7616-1554]{E.~Aakvaag}$^\textrm{\scriptsize 17}$,
\AtlasOrcid[0000-0002-5888-2734]{B.~Abbott}$^\textrm{\scriptsize 121}$,
\AtlasOrcid[0000-0002-0287-5869]{S.~Abdelhameed}$^\textrm{\scriptsize 83b}$,
\AtlasOrcid[0000-0002-1002-1652]{K.~Abeling}$^\textrm{\scriptsize 54}$,
\AtlasOrcid[0000-0001-5763-2760]{N.J.~Abicht}$^\textrm{\scriptsize 48}$,
\AtlasOrcid[0000-0002-8496-9294]{S.H.~Abidi}$^\textrm{\scriptsize 30}$,
\AtlasOrcid[0009-0003-6578-220X]{M.~Aboelela}$^\textrm{\scriptsize 44}$,
\AtlasOrcid[0000-0002-9987-2292]{A.~Aboulhorma}$^\textrm{\scriptsize 36e}$,
\AtlasOrcid[0000-0001-5329-6640]{H.~Abramowicz}$^\textrm{\scriptsize 154}$
% ...
\bigskip
\\

% ...
$^{17}$Department for Physics and Technology, University of Bergen, Bergen; Norway.\\
% ...
$^{30}$Physics Department, Brookhaven National Laboratory, Upton NY; United States of America.\\
% ...
$^{36}$$^{(a)}$Facult\'e des Sciences Ain Chock, Universit\'e Hassan II de Casablanca;$^{(b)}$Facult\'{e} des Sciences, Universit\'{e} Ibn-Tofail, K\'{e}nitra;$^{(c)}$Facult\'e des Sciences Semlalia, Universit\'e Cadi Ayyad, LPHEA-Marrakech;$^{(d)}$LPMR, Facult\'e des Sciences, Universit\'e Mohamed Premier, Oujda;$^{(e)}$Facult\'e des sciences, Universit\'e Mohammed V, Rabat;$^{(f)}$Institute of Applied Physics, Mohammed VI Polytechnic University, Ben Guerir; Morocco.\\
% ...
$^{44}$Physics Department, Southern Methodist University, Dallas TX; United States of America.\\
% ...
$^{48}$Fakult\"{a}t Physik, Technische Universit{\"a}t Dortmund, Dortmund; Germany.\\
% ...
$^{54}$II. Physikalisches Institut, Georg-August-Universit\"{a}t G\"ottingen, G\"ottingen; Germany.\\
% ...
$^{83}$$^{(a)}$Khalifa University of Science and Technology, Abu Dhabi;$^{(b)}$New York University Abu Dhabi, Abu Dhabi;$^{(c)}$United Arab Emirates University, Al Ain;$^{(d)}$University of Sharjah, Sharjah; United Arab Emirates.\\
% ...
$^{102}$CPPM, Aix-Marseille Universit\'e, CNRS/IN2P3, Marseille; France.\\
% ...
$^{121}$Homer L. Dodge Department of Physics and Astronomy, University of Oklahoma, Norman OK; United States of America.\\
% ...
$^{154}$Raymond and Beverly Sackler School of Physics and Astronomy, Tel Aviv University, Tel Aviv; Israel.\\
% ...
$^{a}$ Also at Affiliated with an institute formerly covered by a cooperation agreement with CERN.\\
$^{b}$ Also at An-Najah National University, Nablus; Palestine.\\
$^{c}$ Also at Borough of Manhattan Community College, City University of New York, New York NY; United States of America.\\
$^{d}$ Also at Center for Interdisciplinary Research and Innovation (CIRI-AUTH), Thessaloniki; Greece.\\
$^{e}$ Also at Centre of Physics of the Universities of Minho and Porto (CF-UM-UP); Portugal.\\
$^{f}$ Also at CERN, Geneva; Switzerland.\\
% ...

\end{flushleft}
